\documentclass[hyper]{andp2012}
\usepackage[english]{babel}

\usepackage{color}
\usepackage{soul}
\newcommand{\sm}[1]{\left( \begin{smallmatrix} #1 \end{smallmatrix} \right)}
\title{Complex scattering as canonical transformation: \\ A semiclassical approach in Fock space}
\author[T. Engl]{Thomas Engl\inst{1}\footnote{Corresponding author\quad E-mail:~\textsf{thomas.engl@physik.uni-regensburg.de}}}
\author[J.\,D. Urbina]{Juan Diego Urbina\inst{1}}
\author[Q. Hummel]{Quirin Hummel\inst{1}}
\author[K. Richter]{Klaus Richter\inst{1}}
\address[1]{Institut f\"ur Theoretische Physik, Universit\"at Regensburg, 93040 Regensburg, Germany}
\shortauthors{T. Engl et al.}
\begin{abstract}
We show that a theory of complex scattering between many-body (Fock) states can be constructed such that its classical limit is a canonical transformation thus encoding quantum interference in the semiclassical form of the associated unitary operator. Based on this idea, we study the different coherent effects expected under different choices of the many-body  states and provide different representations of the associated transition probabilities. In this way, we derive exact relations and representations of the scattering process that can be used to attack timely problems related with Boson Sampling. 
\end{abstract}
\shortabstract
\begin{document}
\maketitle

\section{Linear scattering of photons}

We consider a typical scattering scenario, where a highly coherent many photon state of light is injected through waveguides into a complex array of optical elements, such as \textit{e.g.}~in \cite{hong-ou-mandel,QW4,DQW_Coin1,DQW_twophotons,DQW_disorder}. We further assume that decoherence and dephasing due to losses and /or coupling with uncontrolled degrees of freedom can be neglected. The simulation of such a scattering process between multiparticle photonic (or in general, bosonic) input and output states is a computationally hard problem because it involves, as shown bellow, the calculation of permanents of large matrices. The complexity of this problem is expected to render the problem of sampling the space of matrices with a distribution given by their permanents, the Boson Sampling (BS) problem, also hard. Thus, a quantum optical device that samples for us scattering probabilities between many-body states constitutes a quantum computer that eventually beats any classical computer \cite{BosonSampling} in the BS task, an observation that has attracted enormous attention during the last years \cite{BosonSamplingExp1,BosonSamplingExp2,BosonSamplingExp3,BosonSamplingExp4,Malte_tutorial}.

The physical operation of our scattering device consists of mapping the incoming many-photon states $|{\rm in}\rangle$ into the output states $|{\rm out}\rangle$. By injecting the same incoming state several times and counting the number of times we get $|{\rm out}\rangle$ as output, we will eventually obtain the transition probability
\begin{equation}
P(|{\rm in}\rangle \to |{\rm out}\rangle):=|A(|{\rm in}\rangle \to |{\rm out}\rangle)|^{2}=|\langle{\rm out}|{\rm in}\rangle|^{2},
\end{equation}
and our goal is to study this quantity.

As any other quantum state of the field, the $|{\rm in}\rangle,|{\rm out}\rangle$ states belong to the Hilbert (Fock) space ${\cal H}$ of the system, which consists of all possible linear combinations of  Fock states \cite{NO}
\begin{equation}
|{\bold n}\rangle:=|n_{1},n_{2},\ldots,n_{M}\rangle
\end{equation}
specifying the set of integer occupation numbers $n_{1},\ldots,n_{M}$. An occupation number $n_{i}$ specifies how many photons (bosons) occupy the $i$th single-particle state. The choice of these channels (or orbitals) is a matter of convenience, depending on the particular features of the system. In the scattering problem there are two preferred options to construct the Fock space, namely, by defining occupation numbers specifying how many photons occupy a given single-particle state with either incoming or outgoing boundary conditions in the asymptotic region far away from the scatterer. The operators that create a particle in the case of given incoming boundary conditions are denoted by $\hat{b}^{\dagger}$, and their action on the vacuum state $|{\bf 0}\rangle$ produces Fock states in the incoming modes \cite{NO}:
\begin{equation}
|{\bold n}^{\rm in}\rangle:=|n_{1}^{\rm in},n_{2}^{\rm in},\ldots,n_{M}^{\rm in}\rangle=\prod_{i}\frac{\left(\hat{b}^{\dagger}_{i}\right)^{n_{i}^{\rm in}}}{\sqrt{n_{i}^{\rm in}!}}|{\bf 0}\rangle.
\label{eq:Fock-state_in}
\end{equation}
Any operator acting in ${\cal H}$ can be written as a multilinear combination of the creation operators and their adjoints $\hat{b}$, called annihilation operators. The operator algebra is thus uniquely fixed by the canonical commutation relations
\begin{equation}
[\hat{b}_i^{},\hat{b}_j^{}]=0 {\rm \ \ and \ \ \ }[\hat{b}_i^{},\hat{b}_j^\dagger]=\delta_{ij}^{}.
\label{eq:comm}
\end{equation}
Similarly, the operators $\hat{d}^{\dagger}$ create photons in the single-particle states defined by outgoing boundary conditions, and representing physically photons exiting the scattering region along a given channel. A fundamental observation is that the Fock space can be equally well constructed out of the many-body states defined by specifying occupation numbers in the single-particle outgoing states: 
\begin{equation}
|{\bold n}^{\rm out}\rangle:=|n_{1}^{\rm out},n_{2}^{\rm out},\ldots,n_{M}^{\rm out}\rangle=\prod_{i}\frac{\left(\hat{d}^{\dagger}_{i}\right)^{n_{i}^{\rm out}}}{\sqrt{n_{i}^{\rm out}!}}|{\bf 0}\rangle.
\label{eq:Fock-state_out}
\end{equation}
The relation between incoming and outgoing Fock states is fully determined by a single-particle property, namely, the transition amplitude of the single-particle process
\begin{equation}
\sigma_{ij}=\langle 0,\ldots,0,n_{i}^{\rm out}=1,0,\ldots,0|0,\ldots,0,n_{j}^{\rm in}=1,0,\ldots,0 \rangle,
\end{equation}
which defines the single-particle scattering matrix with entries $\sigma_{i,j}$. By comparison with Eqns.~(\ref{eq:Fock-state_in}) and (\ref{eq:comm}), we find
\begin{equation}
\label{eq:bds}
\hat{d}_{j}=\sum_{i}\sigma_{ji}\hat{b}_{i}
\end{equation}
which then allows us to relate the expansion coefficients $c_{\bf n}^{\rm in}$ and $c_{\bf m}^{\rm out}$, appearing in the "in" and "out" representations of an arbitrary many-body state,
\begin{equation}
|\psi\rangle=\sum_{\bf n}c_{\bf n}^{\rm in} |{\bold n}^{\rm in}\rangle=\sum_{\bf m}c_{\bf m}^{\rm out} |{\bold m}^{\rm out}\rangle,
\end{equation}
through the amplitude
\begin{equation}
\label{eq:AF}
A^{\rm F}({\bf n},{\bf m}):=\langle {\bold m}^{\rm out}|{\bold n}^{\rm in}\rangle.
\end{equation}

So far we have focused on the transformation properties between Fock states, but the same questions can be addressed for other type of many-body states. Consider for example the common eigenstates of the incoming creation operators \cite{NO} in the incoming basis,
\begin{equation}
\hat{b}_{i}^{}|{\boldsymbol \phi}^{\rm in}\rangle=\phi_{i}^{\rm in}|{\boldsymbol \phi}^{\rm in}\rangle,
\end{equation} 
so-called coherent states, which are labeled by a continuous set of complex numbers $\phi_{i}$. Although coherent states are not eigenstates of a commuting set of hermitian operators, they can be experimentally prepared \cite{Wolf:2004}, and in some sense they are the most classical states of the electromagnetic field. Again, it can also be shown that both, in- and out-going coherent states, are an (over)complete basis of the Fock space, and the amplitudes
\begin{equation}
A^{\rm C}({\boldsymbol \chi},{\boldsymbol \phi}):=\langle {\boldsymbol \phi}^{\rm out}|{\boldsymbol \chi}^{\rm in}\rangle
\end{equation}
are the matrix elements of a many-body unitary transformation performing the change of representation from incoming to outgoing coherent states.

The third basis set that we are going to discuss is defined by the common eigenstates $|{\bold q}^{\rm in, out}\rangle$ of the so-called quadrature operators \cite{VogelWelsch200607} that correspond to the quantum operator associated with the observable electric field \cite{Cohen-Tannoudji:1997}
\begin{equation}
\hat{q}_{i}^{\rm in}:=\hat{b}_{i}^{}+\hat{b}_{i}^{\dagger} {\rm \ \ , \ }\hat{q}_{i}^{\rm out}:=\hat{d}_{i}^{}+\hat{d}_{i}^{\dagger}.
\end{equation}
It is easy to show that quadrature eigenstates
\begin{equation}
\hat{q}_{i}^{\rm in, out}|{\bold q}^{\rm in, out}\rangle=q_{i}^{\rm in, out}|{\bold q}^{\rm in, out}\rangle
\end{equation}
are labeled by a continuous set of real variables and that they are normalized (to the Dirac delta), complete and orthogonal. We can again define the corresponding transmission amplitude
\begin{equation}
A^{\rm Q}({\bf q},{\bf Q}):=\langle {\bf Q}^{\rm out}|{\bf q}^{\rm in}\rangle.
\end{equation}

The construction of the transformations between the different basis sets is cumbersome but straightforward and we refer the reader to the references \cite{NO,VogelWelsch200607} for further details. We just note that for a given choice of single-particle orbitals, all operators (number, creation/destruction and quadratures) commute with each other if they correspond to different single-particle states (or index $i$). Therefore the results for a given mode \cite{NO,VogelWelsch200607} are sufficient,
\begin{align}
\langle q|n\rangle=&\frac{{\rm e}^{-\frac{q^{2}}{4}}}{\sqrt{2^nn!\sqrt{2\pi}}}{\rm H}_{n}\left(\frac{q}{\sqrt{2}}\right),&& {\rm \ number \ to \ quadrature,}  \nonumber \\
\langle q|\phi\rangle=&\frac{1}{(2\pi)^{1/4}}{\rm e}^{-\frac{|\phi|^2}{2}-\left(\frac{q}{2}-\phi\right)^2+\frac{\phi^2}{2}},&& {\rm \ coherent \ to \ quadrature,} \label{eq:basi-trafo} \\
\langle n|\phi\rangle=&\frac{1}{\sqrt{n!}}\phi^{n}{\rm e}^{-|\phi|^{2}/2},&& {\rm \ coherent \ to \ number}, \nonumber 
\end{align}
where ${\rm H}_{n}(q)$ is the $n$-th Hermite polynomial.

\section{The Boson Sampling problem}

\subsection{Outline of the problem}

With the formalism presented in the last section, we return to our original problem, namely the explicit calculation of scattering amplitudes between Fock states. Using Eq.~(\ref{eq:AF}) and the definitions in Eqns.~(\ref{eq:Fock-state_in}) and (\ref{eq:Fock-state_out}) we get the exact expression
\begin{equation}
\label{eq:BS}
A^{\rm F}({\bf n},{\bf m}):=\langle {\bf 0}|\left[\prod_{j}\frac{\left(\hat{d}_{j}\right)^{m_{j}^{\rm out}}}{\sqrt{m_{j}^{\rm out}}}\right] \left[\prod_{i}\frac{\left(\hat{b}_{j}^{\dagger}\right)^{n_{i}^{\rm in}}}{\sqrt{n_{i}^{\rm in}}}\right]|{\bf 0}\rangle.
\end{equation}

Our goal is to obtain an expression for $A^{\rm F}$, and eventually for the transition probabilities $|A^{\rm F}|^{2}$, in terms of the single-particle scattering matrix ${\boldsymbol \sigma}$. At first glance, due to the absence of interactions this seems to be an easy task since in the scattering process the total amplitude factorizes in terms of the amplitudes of individual, single-particle processes. However, while this is the case for systems of non-interacting {\it distinguishable} particles, for the case of interest here quantum effects due to {\it indistinguishability} render the calculation of scattering amplitudes a hard problem \cite{BosonSampling,Malte_tutorial}, as it is apparent when trying to calculate the amplitudes by substitution of Eq.~(\ref{eq:bds}) into Eq.~(\ref{eq:BS}) and further using the commutator in Eq.~(\ref{eq:comm}) that the complexity of the result is combinatorial in origin. Since the explicit calculation has been reported elsewhere, we just present the final result 
 \begin{equation}
\label{eq:BSFP}
A^{\rm F}({\bf n},{\bf m})={\rm Perm}~{\bf M}({\boldsymbol \sigma}).
\end{equation}
and refer the reader to \cite{BosonSampling,Malte_tutorial} for further details. Following \cite{BosonSampling,Malte_tutorial}, the transition amplitude obtained by this procedure is given by summing up products of entries of ${\boldsymbol \sigma}$, where each term is actually a permutation of the multi dimensional indices labeling output channels. It can be therefore written in terms of a new matrix ${\bf M}({\boldsymbol \sigma})$ (obtained by repeating the $j$-th row of ${\bf \sigma}$ $n_j$ times and the $j$-th column $m_j$ times\footnote{The precise and primary definition of ${\bf M}$ is
$M_{jk}=\sigma_{d_j({\bf n})d_k({\bf m})}$,
with ${\bf d}({\bf n})$ being the $N$-dimensional vector defined by $d_j({\bf n})=\min\left\{k\in\{1,\ldots,M\}\ :\ n_{k-1}<j\leq n_k\right\}$. 
By permuting columns and rows of ${\bf M}$ one arrives at the definition above.}). The key observation is that we indeed {\it sum} all the terms obtained by permuting the second index of this enlarged matrix, resulting in an object known as {\it permanent}.

In this way, the physical scattering of photons provides a physical device that calculates permanents of large matrices. One only needs to repeatedly measure the output state and the accuracy with which the device calculates (or simulates) the precise value of the associated permanent can be made arbitrarily large by repeating the measurement as many times as needed. In a further step, by randomly changing ${\boldsymbol \sigma}$ the device can be used to sample the space of matrices with a weight given by their permanent. It is this task, the BS problem, where under certain conditions it is expected that the quantum device beats any classical computer \cite{BosonSampling}. 

\subsection{Many-body scattering as canonical transformation}

Small scattering devices calculating permanents can be actually realized, with several examples now available \cite{BosonSamplingExp4}, while preparation techniques that allow for coherent creation of correlated photons beyond $N\simeq30$, where the sampling using the quantum device will beat any classical computer, is presently matter of intense research \cite{BosonSampling}. Realistic scenarios, however, seem to reach severe complications already for $N\simeq 12$. Thus, it seems important to explore whether the fundamental aspects of complex many-body scattering allow for other types of implementation, different from the photonic ones. Here we try to approach this question from a more abstract perspective. 

We only demand the matrix ${\boldsymbol \sigma}$ to be a unitary single-particle scattering matrix, 
\begin{equation}
[{\boldsymbol \sigma}^{-1}]_{i,j}=[{\boldsymbol \sigma}]_{j,i}^{*}.
\end{equation}
This means that the calculation of permanents of large matrices is realized by a device with outcomes given by the amplitudes
\begin{equation}
A^{\rm F}({\bf n},{\bf m}):=\langle {\bf m}'|{\bf n}\rangle.
\end{equation}
We call $|{\bf m}\rangle$ the state specified by occupations ${\bf m}$ in the unprimed basis and $|{\bf m}'\rangle$ the  state \ specified by ${\bf m}$ in the primed basis. The later is constructed out of the operators  
\begin{equation}
\hat{b}'_{j}=\sum_{i}u_{j,i}^{}\hat{b}_{i}^{} {\rm \ \ and \ \ }\left(\hat{b}'_{j}\right)^{\dagger}=\sum_{i}u_{j,i}^{*}\left(\hat{b}_{j}^{}\right)^{\dagger}
\label{eq:bbp}
\end{equation}
for {\it any} unitary matrix ${\bf u}$. Note that for the choice ${\bf u}={\boldsymbol \sigma}$ we recover the scattering version with $\hat{b}'_{j}=\hat{d}_{j}^{}$. 

Finally, a straight forward calculation shows that,
\begin{equation}
[\hat{b}'_{i},\hat{b}'_{j}]=0 {\rm \  \  and \ \  \ } [\hat{b}'_{i},\left(\hat{b}'\right)^{\dagger}_{j}]=\delta_{i j}^{}
\end{equation}
follows from Eq.~(\ref{eq:comm}). The transformations (\ref{eq:bbp}) are linear and canonical, the latter because they do not change the algebraic relations between the basic operators. 

We conclude that the BS problem can be realized by any nontrivial device where transition amplitudes are measured between Fock states defined by two different sets of creation operators, defined with respect two different single-particle basis sets. The physical implementation of BS requires then, first of all, a measurement protocol that provides the many-body transformation between Fock states after a linear canonical transformation.

Note that for the BS problem, the essential property of the scattering device is that it provides permanents, and the possibility of connecting permanents with transition amplitudes is entirely due to the linearity of the single-particle transformation. Any physical system where the mapping ${\hat b} \to {\hat b}'$ is nonlinear, as happens for general (non-linear) unitary transformations
\begin{equation}
{\hat b} \to {\hat b}'=\hat{U}^{\dagger}\hat{b}\hat{U}, {\rm \  \ \ }\hat{U}={\rm e}^{iG(\hat{{\bf b}},\hat{{\bf b}}^{\dagger})}, 
\end{equation}
with a hermitian but non-quadratic generator $G$, defines a quantum device that still calculates transition amplitudes but {\it not} permanents.

Using this broader view, it is in principle possible to implement other processes where the output of a measurement is given by permanents and therefore can be used as a basis for the physical implementation of BS. The quantum optical scenario involving scattering of photon states has some very attractive features, in particular that the many-body output states can be indeed measured at the single-photon level by photocounting, while its main drawback is the difficulty to prepare of photonic Fock states with large total number of photons (the state of the art is $N=6$). Since quantum states of indistinguishable bosonic atoms with macroscopic occupations can be prepared by cooling techniques, cold-atoms alternative offer an interesting possibility for BS. The drawback here is the difficulty of performing tomography of many-body cold atom systems at the single-particle level, but advances in this direction are under way \cite{oberth}. 

Assuming for the moment that the measurement of many-body states in cold atom systems reaches the regime of single-atom precision, we sketch a possible BS scenario for such systems. Consider a system of ultracold atoms in an optical lattice, where the hopping amplitude between adjacent sites is $J$ and the strength of the interparticle interaction is $V$. Assume now that initially (at time $t^{-}$) we have $V \gg J$, and the interaction energy is so large that hopping gets completely suppressed \cite{Mott-insulator_BHM}, the so-called Mott phase. In good approximation the ground state of the many-body system is  a Fock state where the occupations refer to the number of atoms in each site, namely, a Fock state constructed out of single-particle states defined by localized (Wannier) orbitals \cite{Lewenstein:2012}. The "quench" scenario is defined by an abrupt change of parameters (possible by tuning the atoms through a Feshbach resonance \cite{feshbach-resonance_ultracold-atoms,feshbach-resonance_ultracold-atoms2}) at time $t^{+}$, such that we have now $J \gg V$. We are interested in the transition amplitudes between the initial state and the eigenstates of the quenched Hamiltonian, where the later are 
again Fock states but built from delocalized (momentum) single-particle orbitals. The calculation of these transition amplitudes is strictly given by $A^{\rm F}({\bf n},{\bf m})$, with the specific choice for $\bf u$ as the matrix that linearly relates the Wannier and the momentum orbitals. Thus, such a device provides permanents as its output. Furthemore, if the on-site energies in the Mott phase are chosen to be random, the matrix $\bf u$ is itself random, and BS can be fully implemented.

\section{Equivalent representations}

We return now to the question of how a general single-particle linear canonical transformation is reflected in the transformation of the different (Fock, quadrature and coherent) many-body states and how the hardness of calculating permanents gets reflected in the different representations.

\subsection{Coherent states}

The simplest transformation between many-body states after a single-particle canonical transformation is the one for the coherent states, and hence we start with this case. Any coherent state can be constructed out of the vacuum state $|{\bf 0}\rangle$ by the application of the displacement operator \cite{NO}
\begin{equation}
\hat{D}({\boldsymbol \phi},{\boldsymbol \phi}^{*})={\rm e}^{{\boldsymbol \phi}^{\ast}\cdot \hat{{\bf b}}+{\boldsymbol \phi}\cdot \hat{{\bf b}}^{\dagger}}
\end{equation}
as
\begin{equation}
|{\boldsymbol \phi}\rangle=\hat{D}({\boldsymbol \phi},{\boldsymbol \phi}^{*})|{\bf 0}\rangle,
\end{equation}
and similarly for the primed states $|{\boldsymbol \psi}'\rangle$
\begin{equation}
|{\boldsymbol \psi}'\rangle={\rm e}^{{\boldsymbol \psi}^{\ast}\cdot \hat{{\bf b}}'+{\boldsymbol \psi}\cdot \hat{{\bf b}}'^{\dagger}}|{\bf 0}\rangle.
\end{equation}
From this, and the defining relation between primed and unprimed canonical operators in Eq.~(\ref{eq:bbp}), we get 
\begin{equation}
|{\boldsymbol \psi}'\rangle=|{\bf u} {\boldsymbol \psi}\rangle.
\end{equation}
Using again well known properties of the coherent states \cite{NO}, the transition amplitude is given by
\begin{equation}
\label{phipsi}
A^{\rm C}({\boldsymbol \phi},{\boldsymbol \psi})={\rm e}^{-\frac{1}{2}{\boldsymbol \phi}^{\ast}\cdot{\boldsymbol \phi}-\frac{1}{2}{\boldsymbol \psi^{\ast}}\cdot{\boldsymbol \psi}+{\boldsymbol \psi}^{\ast}\cdot {\boldsymbol \sigma}\cdot {\boldsymbol \phi}}.
\end{equation}
This result implies in turn for the corresponding transition probability
\begin{equation}
P^{\rm C}({\boldsymbol \phi},{\boldsymbol \psi}):=|A^{\rm C}({\boldsymbol \phi},{\boldsymbol \psi})|^{2}={\rm e}^{-|{\boldsymbol \psi}-{\boldsymbol \sigma}\cdot {\boldsymbol \phi}|^{2}},
\end{equation}
admitting a straightforward interpretation, very much consistent with the idea that quantum coherent states are the most classical states of light: At the classical level, the canonical transformation simply consists of a linear transformation between the field amplitudes given by ${\boldsymbol \phi} \to {\bf u}\cdot {\boldsymbol \phi}$. The classical probability to obtain the state ${\boldsymbol \psi}$ after a canonical transformation of the state ${\boldsymbol \phi}$ is implemented is nonzero only if ${\boldsymbol \psi}={\bf u}\cdot {\boldsymbol \phi}$. In the quantum case, this sharp peak is smoothed into a Gaussian.

In terms of the scattering scenario, the transition probability between coherent states also agrees with intuition: The probability is strongly peaked around the output state labeled by the classical field amplitude resulting from scattering of the classical input field.

\subsection{Quadrature states}

In the same spirit as in the case of coherent states, the transformation rule for quadrature states can be deduced by the corresponding transformation for the defining canonical operators. In the coherent state case, the canonical pair is ${\hat b},{\hat b}^{\dagger}$, and therefore for the quadrature case we must find the set of canonical conjugate partners of the $\hat{q}$'s. The obvious choice that turns out to do the job, is to define \cite{VogelWelsch200607}
\begin{equation}
\hat{q}_{i}^{}:=\hat{b}_{i}^{}+\hat{b}_{i}^{\dagger} {\rm \ \ , \ \ } \hat{p}_{i}^{}:=-i(\hat{b}_{i}^{}-\hat{b}_{i}^{\dagger}). \nonumber
\end{equation}
As the $q$-quadratures, the $p$-quadratures have a complete, orthogonal and Dirac-normalized common set of eigenstates,
\begin{equation}
\hat{p}_{i}|{\bf p}\rangle=p_{i}|{\bf p}\rangle .
\end{equation}
The analogy with the usual position and momentum operators in particle (first quantized) quantum mechanics is evident after using their definition to obtain \cite{VogelWelsch200607}
\begin{equation}
\langle {\bf q}|{\bf p}\rangle=\frac{{\rm e}^{\frac{i}{2} {\bf q}\cdot{\bf p}}}{(4\pi)^{M/2}}.
\end{equation}
However, it must be stressed that quadrature states do not represent any single-particle property at all. In fact, it can be shown that they do not represent states with a well defined total number of particles, thus making their interpretation as any sort of localization property in real space impossible.

Our goal is again to inter-relate the two quadrature states $|{\bf Q}'\rangle$ and $|{\bf q}\rangle$, defined by 
\begin{eqnarray}
\hat{{\bf q}}|{\bf q}\rangle&=&{\bf q}|{\bf q}\rangle, \\
\hat{{\bf q}}'|{\bf Q}'\rangle&=&{\bf Q}|{\bf Q}'\rangle, \nonumber
\end{eqnarray}
using as input the canonical transformation given by
\begin{equation}
\hat{{\bf q}}+i\hat{{\bf p}}\to \hat{{\bf q}}'+i\hat{{\bf p}}'={\bf u}(\hat{{\bf q}}+i\hat{{\bf p}}).
\end{equation}
This canonical transformation can be solved for $\hat{{\bf q}}'$ simply by taking its hermitian part on both sides using the decomposition
\begin{equation}
{\bf u}={\bf u}^{\rm r}+i{\bf u}^{\rm i}
\end{equation}
into real and imaginary parts. The eigenvalue equation defining $|{\bf Q}'\rangle$ is then found to be
\begin{equation}
\left[i{\bf u}^{\rm i}\cdot\frac{\partial}{\partial{\bf q}}-\frac{1}{2}\left( {\bf u}^{\rm r}\cdot {\bf q}-{\bf Q}\right)\right]\langle {\bf Q}'|{\bf q} \rangle=0.
\end{equation}
This can be solved using a Gaussian ansatz to get 
\begin{equation}
\label{eq:Qq}
\begin{split}
A^{\rm Q}({\boldsymbol q},{\boldsymbol Q}):=&\langle {\bf Q}'|{\bf q} \rangle \\
=&\frac{\exp\left\{-\frac{i}{4}\left(\begin{array}{c} {\bf q} \\ {\bf Q} \end{array}\right)\left(\begin{array}{cc} \left({\bf u}^i\right)^{-1}{\bf u}^r & -\left({\bf u}^i\right)^{-1} \\ -\left[\left({\bf u}^i\right)^T\right]^{-1} & {\bf u}^r\left({\bf u}^i\right)^{-1} \end{array}\right)\left(\begin{array}{c} {\bf q} \\ {\bf Q} \end{array}\right)\right\}}{\sqrt{\det\left[-4\pi i{\bf u}\left({\bf u}^i\right)^T\right]}},
\end{split}
\end{equation}
with similar expressions for the $p$-quadrature states.

Using Eq.~(\ref{eq:Qq}), we obtain an interesting result for the transition probability between quadratures,
\begin{equation}
P^{\rm Q}({\boldsymbol q},{\boldsymbol Q}):=|A^{\rm Q}({\boldsymbol q},{\boldsymbol Q})|^{2}=\frac{1}{\left|\det4\pi{\bf u}\left({\bf u}^i\right)^T\right|}.
\end{equation}
It is {\it fully independent of the initial and final states}. In the scattering scenario this means that the probability to obtain a given configuration after measuring the electric field in the output channels is the same for any input and output configuration. As it is clearly seen in Eq.~(\ref{eq:Qq}), however, the amplitudes themselves are very structured functions of the input and output quadrature states and it is only the associated probabilities that display a flat profile.

\section{Exact representations}

Armed with the results of the last section we can now construct different exact expressions for the transition amplitudes between Fock states, $A^{\rm F}({\bf n},{\bf m})$, that, supplemented with an ensemble of random ${\bf u}$ matrices, provide different representations of BS. Since the transition amplitudes, Eqs.~(\ref{phipsi},\ref{eq:Qq}), in both quadrature and coherent representation are not difficult to evaluate, the complexity of calculating permanents must stem from the transformations between the different basis sets. In the following we will make this connection explicit.

Using the transformation rules, Eq.~(\ref{eq:basi-trafo}), between coherent, quadrature and Fock states we obtain \cite{NO,VogelWelsch200607},
\begin{equation}
\label{eq:AFCO}
\begin{split}
A^{\rm F}({\bf n},{\bf m}):=\langle{\bf m}'|{\bf n}\rangle=\frac{1}{\pi^{2M}}\int d{\boldsymbol \psi}d{\boldsymbol \phi}\langle{\bf m}'|{\boldsymbol \psi}'\rangle \langle{\boldsymbol \psi}'|{\boldsymbol \phi}\rangle \langle {\boldsymbol \phi}|{\bf n}\rangle& \\
=\int d{\boldsymbol \psi}d{\boldsymbol \phi}\prod_{i}\frac{\psi_{i}^{m_{i}} \left(\phi_{i}^{\ast}\right)^{n_{i}}{\rm e}^{-\frac{1}{2}|\psi_{i}|^{2}-\frac{1}{2}|\phi_{i}|^{2}}}{\pi^2\sqrt{m_i!n_i!}}A^{\rm C}({\boldsymbol \phi},{\boldsymbol \psi})&,
\end{split}
\end{equation}
and
\begin{equation}
\label{eq:AFQU}
\begin{split}
A^{\rm F}&({\bf n},{\bf m}):=\langle{\bf m}'|{\bf n}\rangle=\int d{\boldsymbol Q}d{\boldsymbol q}\langle{\bf m}'|{\boldsymbol Q}'\rangle \langle{\boldsymbol Q}'|{\boldsymbol q}\rangle \langle {\boldsymbol q}|{\bf n}\rangle \\
&=\int d{\boldsymbol Q}d{\boldsymbol q}\prod_{i}\frac{{\rm H}_{m_{i}}\left(\frac{Q_{i}}{\sqrt{2}}\right) {\rm H}_{n_{i}}\left(\frac{q_{i}}{\sqrt{2}}\right) {\rm e}^{-\frac{Q_{i}^{2}}{4}-\frac{q_{i}^{2}}{4}}}{\sqrt{2^{n_i+m_i+1}\pi n_i!m_i!}}A^{\rm Q}({\boldsymbol q},{\boldsymbol Q}).
\end{split}
\end{equation}
Equations~(\ref{eq:AFCO}) and(\ref{eq:AFQU}) are two equivalent representations of the scattering amplitudes and provide a basis for realizing BS when supplemented with a physical ensemble of unitary matrices ${\bf u}$. The first expression in terms of the coherent state transition amplitude $A^{\rm C}$ is convenient for exact calculations, while the second equation in terms of $A^{\rm Q}$ will be important when we connect BS with a three-step canonical transformation in order to understand its asymptotics for large $N$.

\section{A generating function for transition amplitudes}

It is instructive to show how one finds yet another version of the transition amplitudes using the equivalence of the two representations in Eqns.~(\ref{eq:AFCO}) and (\ref{eq:AFQU}). To this end we use the identities
\begin{equation}
\phi^{n}=\left.\frac{\partial^{n}}{\partial k^{n}}{\rm e}^{k \phi}\right|_{k=0} {\rm \  \ , \ \ }{\rm H}_{n}(q)= {\rm e}^{q^{2}}\left.\frac{\partial^{n}}{\partial k^{n}}{\rm e}^{-(q-k)^{2}}\right|_{k=0}, \nonumber
\end{equation}
which allow us to perform exactly the integrals over the intermediate variables ${\boldsymbol \psi},{\boldsymbol \phi}$ in the coherent state representation and ${\boldsymbol Q},{\boldsymbol q}$ in the quadrature case. After some calculations we get the exact, surprisingly simple expression,
\begin{equation}
\label{eq:main}
\begin{split}
A^{\rm F}({\bf n},{\bf m})=&\left. \left(\prod_{i}\frac{1}{\sqrt{m_{i}!n_{i}!}}\frac{\partial^{m_{i}}}{\partial x_{i}^{m_{i}}} 
\frac{\partial^{n_{i}}}{\partial y_{i}^{n_{i}}}\right){\rm e}^{{\bf x} \cdot {\boldsymbol u} \cdot {\bf y}}\right|_{{\bf x}={\bf y}={\bf 0}} \\
=&\left(\prod_{i}\frac{\sqrt{m_{i}!n_{i}!}}{(-4\pi^{2})}\right)\oint\left(\prod_{i}\frac{dx_{i}dy_{i}}{x_{i}^{m_{i}+1}y_{i}^{n_{i}+1}}\right){\rm e}^{{\bf x} \cdot {\boldsymbol u} \cdot {\bf y}}
\end{split}
\end{equation}
which is one of our main results. It is a generating function providing the transition amplitudes as high-order derivatives of an multivariate exponential function and generalizes \cite{Fyodorov}. Here it is clear that in any representation the complexity of many-body scattering comes from the combinatorics involved in taking high-order derivatives of large products of exponentials. The second expression, obtained by using the Cauchy integral formula (the closed integration contours enclose the origin), further transforms the problem in a way suitable for asymptotic analysis.

The generating function approach provides a way to eventually address some open questions, in particular the calculation of high order moments of the distribution of transition amplitudes (or transition probabilities) over the ensemble of single-particle canonical transformations \cite{BosonSampling}. The particular advantage of this representation is that the average over the unitary group of matrices ${\bf u}$ representing the single-particle canonical transformation can be performed exactly. In section~(\ref{sec:EPQ}) we show how to follow this program in the simpler case of Ginibre (complex) matrices, and provide for the first time exact, explicit expressions for the third moments of the distribution of squared permanents.

So far, all equivalent versions of the transition amplitudes have been obtained by exact, identical transformations. In the rest of this section we will focus on the particular regime of high densities, i.e, when $N:=\sum_{i}n_{i} \gg M$, where we can safely assume that the majority of configurations satisfy
\begin{equation}
\label{eq:HD}
n_{i} \gg 1, m_{i} \gg 1,
\end{equation} 
and powerful methods of asymptotic analysis can be safely applied. However, although BS involves the regime of large $N$ and $M$, it is expected to be a hard problem only in a specific asymptotic limit given by $M \gg N^{2}$ \cite{BosonSampling}, and the high-density limit cannot be used to make statements about it. The study of the behavior of the scattering amplitudes in the appropriate dilute limit of interest for BS is currently under progress.  

If the conditions in Eq.~(\ref{eq:HD}) hold, we can then evaluate the contour integrals in Eq.~(\ref{eq:main}) by the method of steepest descent applied to
\begin{equation}
\begin{split}
&A^{\rm F}({\bf n},{\bf m})=\left(\prod_{i}\frac{\sqrt{m_{i}!n_{i}!}}{-4\pi^{2}}\right) \\
&\times\oint\left(\prod_{i}dx_{i}dy_{i}\right){\rm e}^{-\sum_{i}(n_{i}+1)\log x_{i}-\sum_{i}(m_{i}+1)\log y_{i}+{\bf x} \cdot {\boldsymbol u} \cdot {\bf y}},
\end{split}
\end{equation}
thus making contact with the theory developed in \cite{Str}. Here we are not interested in the technical details of the full calculation of the large-$N$ asymptotics, but instead in the physical interpretation of the saddle point conditions
\begin{eqnarray}
\frac{\partial}{\partial x_{l}}\left[-\sum_{i}n_{i}\log x_{i}-\sum_{i}m_{i}\log y_{i}+{\bf x} \cdot {\bf u} \cdot {\bf y}\right]&=&0, \\
\frac{\partial}{\partial y_{l}}\left[-\sum_{i}n_{i}\log x_{i}-\sum_{i}m_{i}\log y_{i}+{\bf x} \cdot {\bf u} \cdot {\bf y}\right]&=&0
\end{eqnarray}
selecting the optimal values of the, so far, purely formal complex variables ${\bf x},{\bf y}$. Under the variable transformation
\begin{equation}
x_{i}=\sqrt{n_{i}}{\rm e}^{-i\theta_{i}},y_{i}=\sqrt{m_{i}}{\rm e}^{i\chi_{i}}
\end{equation} 
the resulting set of $2M$ complex equations can be reduced to find the $M$ real angles $\chi_{l}$ satisfying the conditions
\begin{equation}
\sum_{l,l'}u_{il}u_{il'}\sqrt{m_{l}m_{l'}}{\rm e}^{i(\chi_{l}-\chi_{l'})}= n_{i}.
\end{equation}
In other words, the asymptotic limit of many-body transition amplitudes for large densities is dominated by configurations $x,y$ satisfying 
\begin{equation}
\label{eq:Shoot}
{\bf y}={\bf u} \cdot {\bf x} {\rm \ \ with \ \ }|y_{l}|^{2}=m_{l} {\rm \ and \ }|x_{i}|^{2}=n_{i}.
\end{equation}
This shows that in the limit of large densities, the calculation of transition amplitudes requires the solution of (\ref{eq:Shoot}), namely the calculation of the phases of the classical input and output field amplitudes (linearly related through ${\bf u}$) required to satisfy {\it shooting} (instead of initial-value) boundary conditions. This interpretation  can be made even more explicit by considering the quadrature representation of the amplitudes. To this end, we consider the chain 
\begin{equation}
\label{eq:chain}
({\bf n},{\boldsymbol \theta}) \to ({\bf q},{\bf p}) \to ({\bf Q},{\bf P}) \to ({\bf N},{\boldsymbol \Theta})
\end{equation}
defined by 
\begin{equation}
q_{i}+ip_{i}=\sqrt{n_{i}}{\rm e}^{i\theta_{i}} {\rm \ \ , \ \ } Q_{i}+iP_{i}=\sqrt{N_{i}}{\rm e}^{i\Theta_{i}}
\end{equation}
and
\begin{equation}
{\bf Q}+i{\bf P}={\bf u}({\bf q}+i{\bf p}).
\end{equation}
The semiclassical approximation for the amplitudes that define the unitary operators representing the first and last canonical transformations of the chain in Eq.~(\ref{eq:chain}),
\begin{equation}
A^{\rm qn}({\bf n},{\bf q})=\langle {\bf q}|{\bf n}\rangle {\rm \ \ , \ \ } A^{\rm QN}({\bf N},{\bf Q})=\langle {\bf Q}|{\bf N}\rangle,
\end{equation}
is given by \cite{canonical_transformation_semiclassics1,canonical_transformation_semiclassics2}
\begin{equation}
\begin{split}
A^{\rm qn}({\bf n},{\bf q})&\simeq \prod_{i=1}^{N}\sqrt{\frac{1}{2\pi i}\frac{\partial^{2} f({n_{i},q_{i}})}{\partial n_{i}\partial q_{i}}}{\rm e}^{if(n_{i},q_{i})}, \\
A^{\rm QN}({\bf N},{\bf Q})&\simeq \prod_{i=1}^{N}\sqrt{\frac{1}{2\pi i}\frac{\partial^{2} F({N_{i},Q_{i}})}{\partial N_{i}\partial Q_{i}}}{\rm e}^{iF(N_{i},Q_{i})}
\end{split}
\end{equation}
in terms of generating functions $f(n,q),F(N,Q)=f(N,Q)$ satisfying
\begin{equation}
\theta=\frac{\partial}{\partial n}f(n,q), {\rm \ \ \ }\Theta=\frac{\partial}{\partial N}f(N,Q).
\end{equation}
Finding these generating functions is a standard problem with explicit solution
\begin{equation}
f(n,q)=\frac{q}{4}\sqrt{4n-q^2}-n\arccos\left(\frac{q}{2\sqrt{n}}\right).
\end{equation}
Interestingly, and contrary to $A^{\rm qn}$ and $A^{\rm QN}$, due to the linearity of the transformation $({\bf q},{\bf p}) \to ({\bf Q},{\bf P})$ the intermediate step $({\bf q},{\bf p}) \to ({\bf Q},{\bf P})$ (responsible for the change in  single-particle representation) is not only approximated but it is in fact exactly given by the semiclassical expression. The result is then identical to $A^{\rm Q}({\bf q},{\bf Q})$ in Eq.~(\ref{eq:Qq}).

We can now construct the semiclassical approximation for the full transformation $({\bf n},{\boldsymbol \theta}) \to ({\bf N},{\boldsymbol \Theta})$ by operator multiplication of the three intermediate transformations,
\begin{equation}
A^{\rm F}({\bf n},{\bf N})=\int d{\bf q}d{\bf Q}\left(A^{\rm QN}({\bf N},{\bf Q})\right)^{\ast}A^{\rm Q}({\bf Q},{\bf q})A^{\rm qn}({\bf n},{\bf q}),
\end{equation}
to get
\begin{equation}
\label{eq:main2}
\begin{split}
A&^{\rm F}({\bf n},{\bf N})= \\
&\begin{split}
\int d{\bf q}d{\bf Q}\frac{\exp\left\{-\frac{i}{4}\left(\begin{array}{c} {\bf q} \\ {\bf Q} \end{array}\right)\left(\begin{array}{cc} \left({\boldsymbol \sigma}^i\right)^{-1}{\boldsymbol \sigma}^r & -\left({\boldsymbol \sigma}^i\right)^{-1} \\ -\left[\left({\boldsymbol \sigma}^i\right)^T\right]^{-1} & {\boldsymbol \sigma}^r\left({\boldsymbol \sigma}^i\right)^{-1} \end{array}\right)\left(\begin{array}{c} {\bf q} \\ {\bf Q} \end{array}\right)\right\}}{\sqrt{\det\left[-4\pi i{\boldsymbol \sigma}^i\left({\boldsymbol \sigma}^i\right)^T\right]}} \\
\times\prod\limits_{j}\frac{\exp\left\{i\left[-\frac{Q_j}{4}\sqrt{4N_j-Q_j^2}+N_j\arccos\left(\frac{Q_j}{2\sqrt{N_j}}\right)\right]\right\}}{\sqrt{2\pi i\sqrt{N_j-Q_j^2/4}}} \\
\times\prod\limits_{j}\frac{\exp\left\{i\left[\frac{q_j}{4}\sqrt{4n_j-q_j^2}-n_j\arccos\left(\frac{q_j}{2\sqrt{n_j}}\right)\right]\right\}}{\sqrt{2\pi i\sqrt{n_j-q_j^2/4}}}.
\end{split}
\end{split}
\end{equation}
This is exactly the same result we obtain by considering the large $n$ limit of the exact representation, Eq.~(\ref{eq:AFQU}), by using the asymptotics
\begin{equation}
\begin{split}
H_{n}(q)\simeq&
\sqrt{\frac{2^{n+1}n^n{\rm e}^{-n+q^2}}{\sqrt{1-\frac{q^2}{2n+1}}}} \\
&\begin{split}
\times\cos\left[\left(n+\frac{1}{2}\right)\arcsin\left(\frac{q}{\sqrt{2n+1}}\right)+\frac{q}{2}\sqrt{2n+1-q^2}\right. \\
\left.-\frac{\pi}{2}\left(n+\frac{1}{2}\right)\right].
\end{split}
\end{split}
\end{equation}
Note that the complexity of many-body scattering is reflected in the coherent sums over quantum mechanical amplitudes explicitly appearing in Eq.~(\ref{eq:BSFP}), namely, quantum interference results in the highly irregular pattern one obtains for the transition probabilities as a function of the incoming and output states \cite{Malte_tutorial}. Very much opposite to the {\it semiclassical} method presented here, {\it quasiclassical} approaches, based on adding probabilities instead of amplitudes, capture only the gross features of these patterns. To stress this point, it is important to understand where quantum interference is hidden in our semiclassical approach. In terms of Eq.~(\ref{eq:BSFP}), by expanding the permanents of ${\bf M}({\boldsymbol \sigma})$ as sums over products of single-particle scattering matrices, these coherent sums over products of single-particle paths can be made very explicit, as in \cite{us}.
 
The semiclassical interpretation of many-body scattering (at least for the case of large occupations) allows us to understand the complexity of the problem, and the origin of massive quantum interference in terms of classical canonical transformations. To this end, consider now the unique canonical transformation implementing the full change of canonical variables $({\bf n},{\boldsymbol \theta}) \to ({\bf N},{\boldsymbol \Theta})$ without the intermediate steps in terms of quadratures. Then the semiclassical theory of quantum canonical transformations indicates that we must find the generating function $w({\bf n},{\bf N})$ that, together with the definitions
\begin{eqnarray}
{\boldsymbol \theta}=\frac{\partial}{\partial {\bf n}}w({\bf n},{\bf N}) &{\rm \ \ , \ \ }& {\boldsymbol \Theta}=\frac{\partial}{\partial {\bf N}}w({\bf n},{\bf N}), \\
\sqrt{N_{i}}{\rm e}^{i\Theta_{i}}&=&\sum_{j}u_{ij} \sqrt{n_{j}}{\rm e}^{i\theta_{j}}, 
\end{eqnarray}
gives the explicit form of the transformation as 
\begin{equation}
{\bf N}={\bf N}({\bf n},{\boldsymbol \theta}) {\rm \ \ , \ \ }{\boldsymbol \Theta}= {\boldsymbol \Theta}({\bf n},{\boldsymbol \theta}),
\end{equation}
in order to write
\begin{equation}
A^{\rm F}({\bf n},{\bf N}) \propto \left|\det\frac{\partial^{2}w({\bf n},{\bf N})}{\partial {\bf n} \partial {\bf N}}\right|^{\frac{1}{2}}{\rm e}^{iw({\bf n},{\bf N})}.
\end{equation}
However, in this case we encounter a new issue that was not present in the canonical transformations we have seen before: although the {\it initial value problem} of finding $({\bf N},{\boldsymbol \Theta})$ from $({\bf n},{\boldsymbol \theta})$ admits a unique solution (given by the transformation equations), the {\it boundary problem} of finding $({\boldsymbol \theta},{\boldsymbol \Theta})$
for given $({\bf n},{\bf N})$ admits a very large set of solutions. Each of these solutions represents a branch $\gamma$ of the multi-valued generating function $w$, and the correct form of the semiclassical approximation to the transition amplitude is then,
 \begin{equation}
A^{\rm F}({\bf n},{\bf N})=\sum_{\gamma} \left|\det\frac{1}{2\pi}\frac{\partial^{2}w_{\gamma}({\bf n},{\bf N})}{\partial {\bf n} \partial {\bf N}}\right|^{\frac{1}{2}}{\rm e}^{iw_{\gamma}({\bf n},{\bf N})+i\mu_{\gamma}\frac{\pi}{4}}.
\end{equation}
Here the index $\mu_{\gamma}$ is a topological property of the particular branch that can be computed from the classical transformation. As expected, this is also the solution of the calculation of the amplitudes using the generating function (\ref{eq:main2}), within the saddle point approximation. Hence the semiclassical origin of both, the complexity of many-body scattering and the massive quantum interference associated with it, is the highly non-linear form (and therefore the multi-valuedness) of the boundary problem connecting occupations.

\section{Distribution of permanents}
\label{sec:EPQ}

In this section we will calculate the first three moments of the distribution of permanents over the (complex) Ginibre ensemble to show exemplary how the representation~(\ref{eq:main}) leads to a solvable combinatorial problem. The calculation is exact, in that it does not involve any asymptotics. It would be of course important to perform a similar calculation in the regime of interest for BS, and this is work under progress.
 
Let $\sigma^2$ denote the variance of the independent real parts and imaginary parts of all matrix elements in ${\bf A}$ and let $N$ be its dimension. We start with an exact representation obtained from (\ref{eq:main}), in a slightly different form
\begin{equation}
{\rm Perm}~{\bf A}=\left.\left(\prod_{i=1}^{N}\frac{\partial^{2}}{\partial x_{i} \partial y_{i}}\right){\rm e}^{{\bf x}^{\tau}{\bf A}{\bf y}}\right|_{{\bf x}={\bf y}={\bf 0}} \,,
\end{equation}
where ${\bf x}=(x_{1},\ldots,x_{N}$) (and similarly for ${\bf y}$) is a column vector and $\tau$ is transposition. This representation allows the Gaussian average to be performed exactly. Define the tensor 
\begin{equation}
{\bf \rho}^{(k)}=\left({\bf y}^{(k)}\right) \left({\bf x}^{(k)}\right)^{\tau}
\end{equation}
such that
\begin{equation}
\sum_{k=1}^{2n}\left({\bf x}^{(k)}\right)^{\tau}{\bf A}\left({\bf y}^{(k)}\right)={\rm Tr} \left[{\bf A}\sum_{k=1}^{2n}{\bf \rho}^{(k)}\right] \,.
\end{equation}
The average of $|{\rm Perm}~{\bf A}|^{2n}$ is evaluated by separating real and imaginary parts of the matrix ${\bf A}$ to get
\begin{equation}
  \begin{split}
		\langle|{\rm Perm}~{\bf A}|^{2n}\rangle=\left(\prod_{k=1}^{n}\prod_{i=1}^{N}\frac{\partial^{2}}{\partial x_{i}^{(2k-1)}\partial y_{i}^{(2k-1)}} \frac{\partial^{2}}{\partial x_{i}^{(2k)}\partial y_{i}^{(2k)}}\right) \\
		\times \left. {\rm e}^{2\sigma^{2}\sum_{i,j=1}^{N}\sum_{k,l=1}^{n}x_{i}^{(2k-1)} y_{j}^{(2k-1)} x_{i}^{(2l)} y_{j}^{(2l)}} \right|_{{\bf x}={\bf y}={\bf 0}} \,,
	\end{split}
\end{equation}
which is equivalent to
\begin{equation}
\label{eq:PerMcomp}
	\begin{split}
		\langle |{\rm Perm}~{\bf A}|^{2n}\rangle={\rm coefficient \ \ of~}\prod_{k=1}^{2n}\prod_{i=1}^{N}x_{i}^{(k)}y_{i}^{(k)} {\rm \ \ in~} \\ \prod_{i,j}\prod_{k,l}\left(1+2\sigma^{2}x_{i}^{(2k-1)}y_{j}^{(2k-1)} x_{i}^{(l)}y_{j}^{(l)}\right).
	\end{split}
\end{equation}
The evaluation of the coefficients in~(\ref{eq:PerMcomp}) is related to the following combinatorial problem.
First of all we can remove the factor $2 \sigma^2$ in~(\ref{eq:PerMcomp}) and in return eventually multiply the overall coefficient with $(2\sigma^2)^{nN}$.
For each value of $i=1,\ldots,N$ all the $x_i^{(k)}, \ k=1,\ldots,2n$, have to appear exactly once.
They come in pairs $x_i^{(k')} x_i^{(l')}$ with $k'$ odd and $l'$ even.
We start by counting the number of ways to combine different factors in $\prod_{k,l=1}^n (1 + x_i^{(2k-1)} x_i^{(2l)})$ to get each variable (for fixed $i$) exactly once.
This is equivalent to counting pairings between $n$ (representing the even indexes) and $n$ (representing the odd indexes), which itself is equivalent to counting permutations of $n$.
We write $ \sm{1&2&\cdots&n \\ P(1)&P(2)&\cdots&P(n)} $ or abreviated $ \sm{P(1)&P(2)&\cdots&P(n)} $ to adress a specific permutation $P \in S_n$.
Specific pairs shall be denoted by the corresponding column $ \sm{k \\ l} = \sm{k \\ P(k)} $.
For all $x$-variables one has to count $N$ independent permutations of $n$.
Writing those one below the other will be referred to as {\it table}.

For the $y$-variables again $N$ permutations of $n$ have to be counted.
Since they come in combination with the $x$-variables in~(\ref{eq:PerMcomp}) they are not independent from the $x$-pairings.
Each tuple $(i,k,l)$ representing a pair in the $N$ $x$-pairings actually comes with a fourth entry as a four-tuple $(i,j,k,l)$.
This means that the pairs $(k,l)$ building up the $y$-pairings have to be taken from the $x$-pairings.
In other words the $y$-pairings have to be a rearrangement of the $x$-pairings, keeping the $(k,l)$-indexes of all pairs.
We will refer to this as a {\it vertical} rearrangement or permutation, depending on the context.
In the process of rearranging identical pairs have to be taken distinguishable (\textit{e.g.}~vertically swapping two identical pairs in the $y$-table has to be counted additionally) since the set of touples $\{(i_1,j_1,k,l),(i_2,j_2,k,l)\}$ is different from the set $\{(i_1,j_2,k,l),(i_2,j_1,k,l)\}$ (if $i_1\neq i_2, j_1 \neq j_2$) although the $(k,l)$-indexes of the two pairs involved are the same.

(i) $n=1$:
There is trivially only one permutation for each $i$ concerning the $x$-variables.
The same holds for the $y$-variables but there are $N!$ ways to vertically rearrange all the $\sm{1 \\ 1}$-pairs.
We get
\begin{equation}
	\langle |{\rm Perm}~{\bf A}|^{2}\rangle = (2\sigma^2)^{N} N! \,.
\end{equation}

(ii) $n=2$:
The two different permutations of \(n=2\) are \( P_1 = \sm{1&2 \\ 1&2} \)
and \( P_2 = \sm{1 & 2 \\ 2 & 1} \), which are {\it incompatible}, meaning they do not share any pair.
Let \(N_1 (M_1)\) and \(N_2 (M_2) \) denote the multiplicities of \(P_1\) and \(P_2\) in the \(x (y)\)-table.
The incompatibility implies $ M_1=N_1, M_2=N_2 $.
The number of ways to distribute these permutations on \(N\) twice is \( \big( \frac{N!}{N_1! N_2!} \big)^2 \) and the number vertical permutations of pairs is \( (N_1!)^2 (N_2!)^2 \).
We get
\begin{eqnarray}
	\langle |{\rm Perm}~{\bf A}|^{4}\rangle &=& (2\sigma^2)^{2N} \sum_{N_1,N_2=0}^N \, \delta_{\sum N_a,N} (N!)^2 \nonumber \\
	&=& (2\sigma^2)^{2N} N! (N+1)! \,.
\end{eqnarray}

(iii) $n=3$:
The \(3!=6\) permutations of \(n=3\) are \( (P_1,\ldots,P_6) = ( \sm{1&2&3}, \sm{2&1&3}, \sm{1&3&2}, \sm{3&2&1}\), \(\sm{2&3&1}, \sm{3&1&2} ) \).
Again we let \(N_a\) and \(M_a\) (\(a=1,\ldots,9\)) denote the multiplicities of the permutations \(P_a\) in the \(x\)- and \(y\)-table respectively.
We define the \(9\) pair-counters \(p_\alpha\)
\begin{eqnarray}
	\label{eq:paircountersC}
	p_1 &=& N_1+N_3 \,, \quad
	p_2 = N_2+N_5 \,, \quad
	p_3 = N_4+N_6 \,, \nonumber \\
	p_4 &=& N_2+N_6 \,, \quad
	p_5 = N_1+N_4 \,, \quad
	p_6 = N_3+N_5 \,, \\
	p_7 &=& N_4+N_5 \,, \quad
	p_8 = N_3+N_6 \,, \quad
	p_9 = N_1+N_2 \nonumber
\end{eqnarray}
for the pairs \( \sm{1\\1}, \sm{1\\2}, \sm{1\\3}, \sm{2\\1}, \sm{2\\2}, \sm{2\\3}, \sm{3\\1}, \sm{3\\2}, \sm{3\\3} \) (in that order).
Taking into account (a) the multinomials for the distributions of the permutations among the \(N\) rows for both \(x\) and \(y\), (b) the restriction to \(y\)-tables that are vertical rearrangements of the \(x\)-table and (c) the vertical permutation of identical pairs for \(y\) yields
\begin{equation}
	\label{eq:PerMcomplex6}
	\begin{split}
		\langle |{\rm Perm}~{\bf A}|&^{6}\rangle = \\
		&\begin{split}
			(2\sigma^2)^{3N}
				\prod_{a=1}^{6} \left( \sum_{N_{a}=0}^N \right) \, \delta_{\sum N_{a},N}
				\prod_{a=1}^{6} \left( \sum_{M_{a}=0}^N \right) \, \delta_{\sum M_{a},N} \\
			\times \prod_{\alpha=1}^{9} \delta_{p_\alpha({\bf N}), p_\alpha({\bf M})} \; \frac{(N!)^2}{\prod_{a} (N_{a}! M_{a}!)} \prod_{\alpha} p_\alpha({\bf M})! \,.
		\end{split}
	\end{split}
\end{equation}
The \(9 \times 6\)-matrix \( \left( \frac{\partial p_\alpha({\bf N})}{\partial N_a} \right)_{\alpha,a} \) has rank \(5\) so there are \(5\) independent restrictions from \( \prod_\alpha \delta_{p_\alpha({\bf N}),p_\alpha({\bf M})} \).
Also the restriction \(\sum_a M_a = N\) is contained when \(\sum_a N_a = N\) applies.
Thus~(\ref{eq:PerMcomplex6}) can also be expressed containing only \(6\) sums.
In the following form the number of sums is reduced to \(7\), keeping one restriction and \(M_1\) independent.
\begin{equation}
	\label{eq:PerMcomplex6reduced}
	\begin{split}
		\langle |{\rm Perm}~{\bf A}|^{6}\rangle = (2\sigma^2)^{3N} (N!)^2
			\prod_{a=1}^{6} \left( \sum_{N_{a}=0}^N \right) \, \delta_{\sum N_{a},N}
			\sum_{M_1=0}^N  \\
		\times \frac{\prod_{\alpha} p_\alpha({\bf N})!}{M_1! \prod_{a} N_{a}! \prod_{a=2}^6 M_a({\bf N},M_1)!}  \,,
	\end{split}
\end{equation}
where the \(M_a\) (\(a>1\)) are given by
\begin{eqnarray}
	M_2 &=& N_1+N_2-M_1 \,, \quad
	M_3 = N_1+N_3-M_1 \,, \nonumber \\
	M_4 &=& N_1+N_4-M_1 \,, \quad
	M_5 = N_5-N_1+M_1 \,, \\
	M_6 &=& N_6-N_1+M_1 \nonumber
\end{eqnarray}
and $ \frac{1}{(-m)!} := 0 $ for $ k \in \mathbb{N}\backslash\{0\} $.
Applying~(\ref{eq:PerMcomplex6reduced}) we evaluate the scaled third moment $\langle |{\rm Perm}~{\bf A}|^{6}\rangle / (2\sigma^2)^{3N} / (N!)^3$ for the lowest $N$ to
$6$, $18$, $\frac{122}{3}$, $79$, $140$, $\frac{10508}{45}$, $\frac{13068}{35}$, $579$, $\frac{276442}{315}$, $\frac{228754}{175}$, $\frac{3697434}{1925}$, $\frac{48374363}{17325}$, $\frac{12084328}{3003}$, $\frac{55026632}{9555}$, $\frac{5536562488}{675675}$, $\frac{290360139}{25025}$, $\frac{3748239326}{229075}$, $\frac{73954590386}{3216213}$, $\frac{156246017726}{4849845}$, $\frac{33081258263}{734825}$, $\frac{95883756128092}{1527701175}$, $\frac{767871070556}{8793675}$, $\frac{750199663660}{6186609}$
for $N=1,\ldots,23$ respectively and use this to estimate an assymptotically exponential (as opposed to factorial) scaling of this quantity proportional to ${\rm e}^{\lambda N} N^\nu (1+\mathcal{O}(\frac{1}{N}))$ with $\lambda \sim 0.3$.

\section{Conclusions}
We have shown that the usual many body scattering scenario realizing the Boson Sampling problem (in the sense of sampling over an ensemble of large matrices using as weight their permanents) is a particular case of a much more general kind of physical situations where the transition amplitudes between many-body Fock states built from two different single-particle basis sets are measured. Within this general scenario, Boson Sampling requires the calculation of the many-body unitary operator representing a linear, canonical transformation at the single-particle level. We have provided different versions of the problem, obtained by expressing this transition amplitudes in different intermediate basis like coherent states and quadrature states of the field. Starting with these exact representations, we performed an asymptotic analysis valid in the limit of large occupations and provide their semiclassical approximation in the spirit of coherent sums over solutions of a classical boundary problem. Along the way, we have derived an exact form of the many body transition amplitudes, equivalent to the calculation of permanents, and use it to derive exact results for the moments of the distribution of permanents over the Ginibre ensemble. Work on the extension of our asymptotic analysis into the regime of low densities, where BS is expected to be hard, is currently under way.  

\begin{acknowledgements}
We thank M.~Tichy, A.~Buchleitner, J.~Kuipers and V.~S.~Shchesnovich for valuable discussions and three anonymous referees for their help in improving the paper.
\end{acknowledgements} 

\bibliographystyle{andp2012}
\bibliography{Engl}

\end{document}